\documentstyle[12pt]{article}

\begin{document}

\begin{titlepage}
\begin{flushright}
\today \\
BA-TH/94-170\\
DPUR 69\\
WU-HEP-94-3\\
\end{flushright}
\vspace{.5cm}
\begin{center}
{\LARGE Exponential behavior of a quantum system
in a macroscopic medium}\\ [.5cm]

{\large Hiromichi NAKAZATO,$^{(1)}$
     Mikio NAMIKI$^{(2)}$ \\
     and Saverio PASCAZIO$^{(3)}$ \\
           \quad    \\
        $^{(1)}$Department of Physics,
University of the Ryukyus \\ Okinawa 903-01, Japan \\
        $^{(2)}$Department of Physics,
Waseda University \\ Tokyo 169, Japan \\
        $^{(3)}$Dipartimento di Fisica, Universit\^^ {a} di Bari \\ and
Istituto Nazionale di Fisica Nucleare, Sezione di Bari \\ I-70126  Bari, Italy

}

\vspace*{.5cm} PACS: 03.80.+r; 03.65.Bz; 05.20.-y; 05.30.-d \vspace*{.5cm}

{\small\bf Abstract}\\ \end{center}

{\small  An exponential behavior at all times is derived for a solvable
dynamical model in the weak-coupling, macroscopic limit. Some implications
for the quantum measurement problem are discussed, in particular
in connection with dissipation. }


\end{titlepage}

\newpage


\setcounter{equation}{0}

`Decoherence'
has become an important keyword in the quantum measurement theory \cite{von}
\cite{Zurek}. Many people ascribe the measurement process to
a dephasing process \cite{Zurek} \cite{MN} and have tried to derive it as
a consequence of the interaction with measuring devices,
within the framework of quantum mechanics.
Because `decoherence' technically means the elimination of the off-diagonal
elements of the density matrix, a system described by such a diagonal density
matrix should exhibit a purely stochastic behavior and we naturally expect
a close connection with a dissipative and irreversible behavior \cite{Leggett}.

On the other hand, the
temporal evolution of a quantum mechanical system, initially
prepared in an eigenstate of the unperturbed Hamiltonian, is known to be
roughly characterized
by three distinct regions \cite{nm}: A Gaussian behavior at short times,
a Breit-Wigner exponential decay at intermediate times, and a
power law at long times.
It is well known that the asymptotic dominance of the exponential behavior is
representative of
a purely stochastic evolution and can be derived
quantum mechanically \cite{vanHove} in the weak-coupling, macroscopic limit
(the so-called van Hove's limit).
One may expect a close connection between dissipation and
exponential decay \cite{Leggett}.
The Gaussian short-time behavior is of particular significance
due, in particular, to the so-called quantum Zeno effect
\cite{Misra}.

The aim of this Letter is to discuss the temporal behavior
of a solvable dynamical model which describes the interaction between a
microscopic and a macroscopic system
in the weak-coupling, macroscopic limit.
Some implications for the quantum measurement problem will also be analyzed,
in particular in connection with dissipation.
We shall base our discussion on the AgBr model \cite{AgBr},
that has played an important role
in the quantum measurement problem, and its
modified version \cite{NaPa3}, which
is able to take into account energy-exchange processes.
Our exact calculation shows that the model realizes the so-called diagonal
singularity \cite{vanHove} and can display
the occurrence of an exponential regime {\em at all times} in the
weak-coupling, macroscopic limit.

The modified AgBr Hamiltonian \cite{NaPa3} describes the interaction
between an
ultrarelativistic particle $Q$ and a 1-dimensional $N$-spin array
($D$-system).
The array is a caricature of a linear ``photographic emulsion" of AgBr
molecules, when one identifies the {\em down} state of the spin with
the undivided molecule and the {\em up} state with the dissociated
molecule (Ag and Br atoms).
The particle and each molecule interact via a spin-flipping local potential.
The total Hamiltonian for the $Q+D$ system reads
\begin{equation}
H = H_{0} + H',  \qquad
\qquad H_0 = H_{Q} + H_{D},
\label{eq:totham}
\end{equation}
where $H_{Q}$ and $H_{D}$,
the free Hamiltonians of the $Q$ particle and of the ``detector" $D$,
respectively, and the interaction Hamiltonian $H'$ are written as
\begin{eqnarray}
H_{Q} & = & c \widehat{p},    \qquad
     H_{D}  =  \frac{1}{2}  \hbar  \omega
  \sum_{n=1}^{N}  \left( 1+\sigma_{3}^{(n)} \right) , \nonumber  \\
H' & = & \sum_{n=1}^{N} V(\widehat{x}- x_n)
  \left[ \sigma_{+}^{(n)} \exp \left( -i \frac{\omega}{c}
  \widehat{x} \right) + \sigma_{-}^{(n)} \exp \left( + i \frac{\omega}{c}
  \widehat{x} \right) \right],
\label{eq:H}
\end{eqnarray}
where $\widehat{p}$ is the momentum of the $Q$ particle, $\widehat{x}$
its position,
$V$ a real potential, $x_n\; (n=1,...,N)$ the positions of the
scatterers in the array $(x_n>x_{n-1})$ and $\sigma_{i,\pm}^{(n)}$
the Pauli matrices acting on the $n$th site.
An interesting feature of the above Hamiltonian, as compared to the
original one \cite{AgBr}, is that we are not neglecting
the energy $H_D$ of the array, namely the energy gap
between the two states of each molecule.
This enables us to take into account energy-exchange
processes between $Q$ and the spin system $D$.
The original Hamiltonian \cite{AgBr} is reobtained in the $\omega=0$ limit.

The temporal evolution of the system under investigation is best
disclosed by studying the behavior of the propagator.
Following van Hove's pioneering work \cite{vanHove}, one could calculate
the propagator perturbatively in this model.
Observe that the interaction Hamiltonian $H'$ has
nonvanishing matrix elements only between those eigenstates of $H_0$
whose spin-quantum numbers differ by one.
It is important to note, in this connection, that the $Q$
particle state $\vert cp\rangle$,
characterized by the energy $cp$, is changed by $H'$ into the state
$\vert cp\pm\hbar\omega\rangle$, if $\omega\not=0$.
We can therefore expect a dissipation effect and the
appearance of the diagonal singularity, which leads to the master equation
\cite{vanHove}.
One could further develop the perturbative treatment just following
Ref.~\cite{vanHove};
however, the solvability of the present model enables us
to perform a nonperturbative
treatment and yields an exact expression for the propagator.

The evolution operator in the interaction picture
\begin{equation}
U(t,t') = e^{iH_{0}t/\hbar} e^{-iH(t-t')/\hbar}
              e^{-iH_{0}t'/\hbar} =
          e^{-i \int_{t'}^{t} H_{I}'(t'') dt''/\hbar },
     \label{eq:evol}
\end{equation}
where $H_{I}'(t)$ is the interaction Hamiltonian in the
interaction picture, can be computed exactly \cite{NaPa3} as
\begin{equation}
U(t,t') = \prod_{n=1}^{N} \exp \left( -\frac{i}{\hbar} \int_{t'}^{t}
   V(\widehat{x} + ct''- x_n) dt''
  \left[ \sigma_{+}^{(n)} \exp \left( -i \frac{\omega}{c}
  \widehat{x} \right) + \mbox{h.c.} \right]  \right) .
\label{eq:solut}
\end{equation}
In what follows, we set $t'=0$ for notational simplicity.
Define
\begin{equation}
\alpha_n \equiv \alpha_n(\widehat{x},t) \equiv
      \int_{0}^{t} V(\widehat{x} + ct'- x_n) dt' /\hbar ,
\label{eq:defalpha}
\end{equation}
which can be viewed as a ^^ ^^ tipping angle" of the $n$th spin
if one identifies $V$
with a magnetic field $B$ \cite{Hiyama}, and
\begin{equation}
\sigma^{(n)}_\pm (\widehat{x}) \equiv
\sigma_{\pm}^{(n)} \exp \left( \mp i \frac{\omega}{c} \widehat{x} \right) ,
 \nonumber
\end{equation}
which satisfy, together with $\sigma^{(n)}_3$, the $SU(2)$ algebra
\begin{equation}
\left[ \sigma^{(n)}_- (\widehat{x}), \sigma^{(n)}_+ (\widehat{x})
\right]=-\sigma^{(n)}_3,
\qquad
\left[ \sigma^{(n)}_\pm (\widehat{x}),- \sigma^{(n)}_3 \right]=
\pm 2 \sigma^{(n)}_\pm (\widehat{x}).
         \label{eq:newalgg}
\end{equation}
We can now return to the Schr\"odinger picture by inverting
eq.~(\ref{eq:evol}).
We disentangle the exponential \cite{Bogol}
in $U$ by making use of eq.~(\ref{eq:newalgg}) and obtain
\begin{equation}
e^{-iHt/\hbar} = e^{-iH_{0}t/\hbar} \prod_{n=1}^N
   \left(   e^{-i \tan(\alpha_n) \sigma^{(n)}_+(\hat{x})}
        e^{-\ln\cos(\alpha_n) \sigma^{(n)}_3}
        e^{-i \tan(\alpha_n) \sigma^{(n)}_-(\hat{x})} \right).
     \label{eq:relinv}
\end{equation}
Notice that the evolution operators (\ref{eq:solut}) and
(\ref{eq:relinv}) are expressed in a factorized form:
This is a property of a rather general class of similar Hamiltonians
\cite{Sun}.

We shall now concentrate our attention on the situation in which the $Q$
particle is initially at the position $x' < x_1$, where $x_1$ is the
position of the first scatterer in the linear array, and is moving towards
the array with speed $c$.
The spin system is initially set in the ground state $|0\rangle_N$ of the
free Hamiltonian $H_D$ (all spins down).
This choice of the ground state is meaningful
from a physical point of view, because the $Q$ particle is initially
outside $D$.

The propagator, defined by
\begin{equation}
G(x,x',t) \equiv \, \langle x| \otimes\, _N\langle 0|
  e^{-iHt/\hbar} |0 \rangle_N \otimes |x' \rangle  ,
     \label{eq:propG}
\end{equation}
is easily calculated from eq.~(\ref{eq:relinv}) and we obtain
\begin{eqnarray}
G(x,x',t) & = & \langle x|\otimes \,
     _N \langle 0| e^{-ic\hat{p}t/\hbar}\prod_{n=1}^N
      \left( e^{-\ln \{ \cos \left[ \alpha_n(\hat{x},t)
       \right] \} \sigma^{(n)}_3} \right)
       |0 \rangle_N \otimes |x' \rangle \nonumber \\
   & = & \langle x|x'+ct \rangle
      \prod_{n=1}^N
      \left( e^{\ln \{ \cos \left[ \alpha_n(x',t) \right]\} } \right)
            \nonumber \\
 & = & \delta (x-x'-ct)
      \prod_{n=1}^N
      \cos \alpha_n(x',t) .
     \label{eq:Gstart}
\end{eqnarray}
Observe that, due to the choice of the free Hamiltonian $H_Q$ in
eq.~(\ref{eq:H}), the $Q$ wave packet does not disperse, and moves
with constant speed $c$.
We place the spin array at the far right of the origin ($x_1>0$) and consider
the case where potential $V$ has a compact support and the $Q$ particle
is initially located at the origin $x'=0$, i.e.\
well outside the potential region of $D$.
The above equation shows that the evolution of $Q$ occurs only along
the path $x=ct$.
Therefore we obtain
\begin{equation}
G(x,0,t) = \delta (x-ct)\prod_{n=1}^N
      \cos \widetilde{\alpha}_n(t) ,\qquad
\widetilde{\alpha}_n(t) \equiv \int_{0}^{ct} V(y - x_n) dy/\hbar c.
        \label{eq:Ginter}
\end{equation}
This result is {\em exact}.

Let $V_{0} \Omega \equiv \int_{-\infty}^{\infty} V(x)dx$
and call the ^^ ^^ spin-flip" probability $q$, i.e.\
the probability of dissociating one AgBr molecule,
\begin{equation}
q \equiv \sin^{2} \widetilde{\alpha}_n (\infty) =
 \sin^{2} \left( \frac{V_{0} \Omega}{\hbar c} \right).
  \label{eq:sfprob}
\end{equation}
Note that $qN\equiv\overline{n}$ represents the average number of
excited molecules.
We shall now consider the weak-coupling, macroscopic limit \cite{NaPa3}
\begin{equation}
q \simeq \left( \frac{V_{0} \Omega}{\hbar c} \right)^2 = O(N^{-1}) ,
 \label{eq:wcmacro}
\end{equation}
which is equivalent to the requirement that $\overline{n}=qN$ be finite.
Notice that if we set
\begin{equation}
x_n = x_1 + (n-1) \Delta,\qquad L=x_N - x_1=(N-1)\Delta,
 \label{eq:xn}
\end{equation}
the scaled variable $z_n\equiv x_n/L$ can be considered
as a continuous one $z$ in the above
limit, for $\Delta/L\to0$ as $N\to\infty$.
Therefore, a summation over $n$ is to be replaced by a definite integration
\begin{equation}
q\sum_{n=1}^Nf(x_n) \rightarrow
q\frac{L}{\Delta}\int_{x_1/L}^{x_N/L}f(Lz)dz
\simeq\overline{n}\int_{x_1/L}^{x_N/L}f(Lz)dz.
    \label{eq:somma}
\end{equation}
This type of integration gives a finite result if the function $f$ is scale
invariant, because the integration volume is considered to be finite from
the physical point of view; in fact, the quantities $x_1/L$ and $x_N/L$ should
be of the order of unity even in the $L\to\infty$ limit.
It will be shown below [eq.~(\ref{eq:vaimo})]
that in the present case the function $f$ is
indeed scale invariant.

For the sake of simplicity, we shall restrict our attention to the case of
$\delta$-shaped potentials, by setting $V(y) = (V_0 \Omega) \delta(y)$.
This hypothesis is in fact too restrictive:
In the following, we shall see that the requirement that $V$ has a compact
support (local potentials) would suffice.
We obtain
\begin{eqnarray}
G & \propto & \exp \left( \sum_{n=1}^N \ln \left\{
      \cos \int_{-x_n}^{ct-x_n} (V_0 \Omega/\hbar c) \delta(y) dy
       \right\} \right)
      \nonumber   \\
   & = & \exp \left( \sum_{n=1}^N \ln \left\{
      \cos \left[ (V_0 \Omega/\hbar c) \theta(ct-x_n) \right] \right\} \right)
      \nonumber   \\
   & \rightarrow & \exp \left( - \frac{\overline{n}}{2}
      \int_{x_1/L}^{x_n/L} \theta(ct-Lz) dz \right)
      \nonumber   \\
   & = & \exp \left( - \frac{\overline{n}}{2} \left[
      \frac{ct-x_1}{L}  \theta(x_N-ct)\theta(ct-x_1)
      + \theta(ct-x_N) \right]  \right),
   \label{eq:vaimo}
\end{eqnarray}
where $\theta$ is the step function and the arrow denotes the
weak-coupling, macroscopic limit (\ref{eq:wcmacro}).

This brings about an exponential regime {\em as soon as the interaction
starts}:
Indeed, if $x_1 < ct < x_N$,
\begin{equation}
G \propto \exp \left( - \overline{n}
      \frac{c(t-t_0)}{2L} \right),
   \label{eq:explaw}
\end{equation}
where $t_0 = x_1/c$ is the time at which the $Q$ particle meets the first
potential.
Notice that there is {\em no} Gaussian behavior at short times and
{\em no} power law at long times.

Observe that if $ct > x_N$ (that corresponds to the case in which $Q$ has
gone through $D$ and the interaction is over) we have
\begin{equation}
G \propto e^{-\overline{n}/2}.
   \label{eq:tinflim}
\end{equation}
This could also be obtained directly from eq.~(\ref{eq:Ginter}),
and is in complete agreement with previous results \cite{NaPa3},
because $|G|^2$ is nothing but the probability that $Q$ goes
through the spin array {\em and} leaves it in the ground state.

It is well known \cite{nm} \cite{FGR} that deviations from exponential
behavior at short times are a consequence of the finiteness of the mean
energy of the initial state. If the position eigenstates in
eq.~(\ref{eq:propG}) are substituted with wave packets of size $a$,
a detailed calculation shows that
the exponential regime is attained a short time after $t_0$,
of the order of $a/c$,
which, in the present model, can be made arbitrarily small.
The region $t\sim t_0+O(a/c)$ may be viewed as a possible residuum of the
short-time Gaussian-like behavior.
For this reason, the temporal behavior derived in this Letter is not in
contradiction with some general theorems
\cite{nm} \cite{Misra} \cite{FGR}.

What causes the occurrence of the exponential behavior
displayed by our model~?
Our analysis suggests that the exponential behavior is mainly due to
the locality of the potentials $V$ and the factorized form  of the evolution
operator $U$.
As already stressed, the factorization is a general property of a class of
similar Hamiltonians \cite{Sun}.

It is very interesting to bring to light the
profound link between the weak-coupling, macroscopic
limit $qN = \overline{n}=$ finite considered in this
Letter and van Hove's ^^ ^^ $\lambda^2 T$" limit \cite{vanHove}.
First, it is important to note that in the weak-coupling, macroscopic limit,
van Hove's so-called diagonal singularity naturally appears in the
present model.
It is easy to check that for each diagonal matrix element of $H'^2$,
there are $N$ intermediate-state contributions:
Indeed, for example
\begin{equation}
\langle0,\ldots,0\vert H'^2\vert0,\ldots,0\rangle
=\sum_{j=1}^N\vert\langle0,\ldots,0\vert H'\vert0,\ldots,0,1_{(j)},0,
                   \ldots,0\rangle\vert^2.
   \label{eq:hsqu}
\end{equation}
On the other hand,
at most 2 states can contribute
to each off-diagonal matrix element of $H'^2$.
This ensures that only the diagonal matrix elements are kept in the
weak-coupling, macroscopic limit, $N\to\infty$ with $qN<\infty$, which is
the realization of diagonal singularity in our model.
The link with the $\lambda^2 T$ limit is easily evinced 
by observing that $q$, in
eq.~(\ref{eq:wcmacro}), is nothing but the square of a coupling
constant (van Hove's $\lambda$),
and that $N (\propto L)$ can be considered proportional to
the total interaction time $T$.
Notice that the ^^ ^^ lattice spacing" $\Delta$, the inverse of which
corresponds to a density in our 1-dimensional model, can be kept finite
in the limit.
(In such a case, we have to express everything in terms of scaled
variables, that is, $\tau\equiv t/L$, $z_1$ and $z_N$, introduced just after
eq.~(\ref{eq:xn}) and $\zeta\equiv a/L$, where $a$ is the size of the wave
packet.)

The role played by the
energy gap $\omega$ need also be clarified.
As a matter of
fact, $\omega$ plays a very important role to guarantee the consistency
of the physical framework:
If $\omega=0$, all spin states would be energetically
degenerated and the choice of
the $\sigma_3$-diagonal representation would be quite {\it arbitrary\/}.
In other words, a nonvanishing
$\omega$ (or $H_D$) logically allows us
to use the eigenstates of $\sigma_3$ in order to evaluate
the relevant matrix elements.
Though $\omega$ does not appear in our final result (\ref{eq:explaw}), it
certainly does in other propagators like $\langle c\tilde p| \otimes\, _N
\langle 0| e^{-iHt/\hbar} |0 \rangle_N \otimes |c\tilde p' \rangle$ where
$|c\tilde p\rangle$ stands for a wave packet.

Finally, let us comment on the connection between dissipation and exponential
decay.
Leggett \cite{Leggett},
by discussing the role of the environment in connection with the collapse of
the wave function,
stressed the central relevance of the problem of
dissipation to the quantum measurement theory.
We find his argument very convincing. However, at the same time, it seems
to us that the idea need be sharpened:
The temporal behavior derived in this Letter,
yielding an exponential ``probability dissipation",
is certainly related to dephasing (``decoherence") effects
of the same kind of those encountered in quantum measurements.
The exchange of energy between the particle and the ``environment"
(our spin system) can be considered practically irreversible.
On the other hand, in the present model, the final result (\ref{eq:explaw})
does {\em not} depend on $\omega$.
We do not want to draw any crucial
conclusion on this issue, starting from the present
analysis, on the basis of
the particular model considered. We leave this as an open problem that
is certainly worth further investigation.



\end{document}